\documentclass{article}

\usepackage{arxiv}
\usepackage[utf8]{inputenc} 
\usepackage[T1]{fontenc}    
\usepackage[nolist]{acronym}
\usepackage{hyperref}   
\usepackage{graphicx}
\usepackage{csquotes}
\usepackage{microtype}
\usepackage{censor}
\usepackage{rotating}
\usepackage{pdflscape}
\usepackage{array}
\usepackage{booktabs}
\usepackage{threeparttable}
\usepackage{threeparttablex}
\usepackage{multirow}
\usepackage{longtable}
\usepackage{xcolor}

\usepackage{doi}
\usepackage[backend=biber,style=authoryear-comp,sorting=ynt]{biblatex}
\usepackage{siunitx}
\usepackage{dcolumn}
\usepackage{rotating} 

\usepackage{wrapfig}
\usepackage{float}
\usepackage{colortbl}
\usepackage{pdflscape}
\usepackage[normalem]{ulem}
\usepackage{multicol}
\usepackage{hyperref}

\usepackage{xspace}

\title{Media Framing Moderates Risk-Benefit Perceptions and Value Tradeoffs in Human-Robot Collaboration}
\renewcommand{\shorttitle}{Media Framing in Human-Robot Collaboration}

\author{
\href{https://orcid.org/0000-0003-2837-5181}{\includegraphics[scale=0.06]{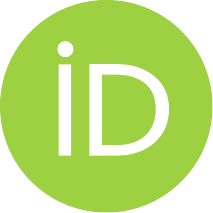}\hspace{1mm}Philipp Brauner} \\
	Communication Science\\
	RWTH Aachen University\\
	52056 Aachen, Germany\\
\And
\href{https://orcid.org/0000-0001-7054-8520}{\includegraphics[scale=0.06]{orcid.pdf}\hspace{1mm}Felix Glawe} \\
	Communication Science\\
	RWTH Aachen University\\
	52056 Aachen, Germany
\And
\href{https://orcid.org/0000-0002-9030-6999}{\includegraphics[scale=0.06]{orcid.pdf}\hspace{1mm}Luisa Vervier} \\
	Communication Science\\
	RWTH Aachen University\\
	52056 Aachen, Germany
\And
\href{https://orcid.org/0000-0002-6105-4729}{\includegraphics[scale=0.06]{orcid.pdf}\hspace{1mm}Martina Ziefle} \\
	Communication Science\\
	RWTH Aachen University\\
	52056 Aachen, Germany}

\begin{document}

\maketitle

\begin{abstract}
Public acceptance of industrial human–robot collaboration (HRC) is shaped by how risks and benefits are perceived by affected employees. Positive or negative media framing may shape and shift how individuals evaluate HRC.
This study examines how message framing moderates the effects of perceived risks and perceived benefits on overall attributed value.
In a pre-registered study participants (N=1150) were randomly assigned to read either a positively or negatively framed newspaper article in one of three industrial contexts (autonomy, employment, safety) about HRC in production.
Subsequently, perceived risks, benefits, and value were measured using reliable and publicly available psychometric scales.
Two multiple regressions (one per framing condition) tested for main and interaction effects.
Framing influenced absolute evaluations of risk, benefits, and value.
In both frames, risks and benefits significantly predicted attributed value.
Under positive framing, only main effects were observed (risks: $\beta \approx -.52$; benefits: $\beta \approx .45$).
Under negative framing, both predictors had stronger main effects (risks: $\beta \approx -.69$; benefits: $\beta \approx .63$) along with a significant negative interaction ($\beta \approx -.32$), indicating that higher perceived risk diminishes the positive effect of perceived benefits.
Model fit was higher for the positive frame ($R^2 = .715$) than for the negative frame ($R^2 = .583$), indicating higher explanation of the variance in value attributions.
Framing shapes the absolute evaluation of HRC and how risks and benefits are cognitively integrated in risk–benefit tradeoffs.
Negative framing produces stronger but interdependent effects, whereas positive framing supports additive evaluations.
These findings highlight the role of strategic communication in fostering acceptance of HRC and underscore the need to consider framing in future HRC research.
\end{abstract}

\keywords{Media Framing, Trust in Robots, 
	Industry 4.0, Risk-Benefit Analysis, Human-Robot Collaboration (HRC), Dual Process Model, Risk communication, Public Perception, Psychometric Evaluation}

\section{Introduction}\label{introduction}

The digital transformation of industrial production \parencite{Brauner2021} or Industry 4.0 \parencite{Kagermann2015} reshapes how manufacturing systems are designed, operated, and experienced \parencite{Liao2017}.
One key aspect of this transformation is the integration of advanced automation through intelligent systems and robotics \parencite{WorldRoboticsIndustrialRobots2024}.
But despite increasing automation, future factories will not be deserted, but equipped with robotic systems that can operate alongside human workers.
Among these technologies, human-robot collaboration (HRC) has emerged as a key enabler of flexible, efficient, and responsive production environments \parencite{Villani2018,Coronado2022}.
HRC systems are designed to leverage the strengths of both humans (e.g., adaptability, decision-making) and robots (e.g., precision, endurance), allowing them to share tasks and physical space in dynamic settings.
This collaborative paradigm marks a shift from traditional automation, which often replaced human labor, to augmentative automation, where the goal is to support and empower human workers.
Yet, one of the grand challenges for robotics is the integration of robots in our (social) work environments and ensuring that future robotic systems are compatible with our social dynamics, moral norms and values \parencite{Yang2018,Coronado2022}.

The success of HRC does not depend solely on technical performance but also on how people feel about production robots, social acceptance, and trust.
A growing body of research highlights that technologies introduced with insufficient consideration of human factors can lead to resistance, misuse, or underutilization \parencite{Parasuraman1997,Hancock2011,Hoff2015,Coronado2022}.
Acceptance is particularly critical in high-risk, high-stakes environments such as production, where perceived risks and benefits are also strong predictors of whether humans will engage with robotic systems \parencite{Murashov2016,deVisser2018}.
A lack of acceptance can result in implementation failures, reduced efficiency, or even ethical and legal challenges \parencite{Wang2020,Liehner2021}
Therefore, ensuring that HRC is not only technically effective but also \emph{perceived} as safe, beneficial, and trustworthy is a critical design and communication objective.
Moreover, acceptance plays a central role in broader organizational goals, as it influences productivity, team cohesion, well-being, and the long-term sustainability of digital transformation initiatives, especially amid the growing shortage of skilled workers in times of demographic change \parencite{Helms2024}.

In this context, \emph{how} information about HRC is communicated becomes relevant.
Research on framing in different domains suggests that focusing on benefits (\emph{positive framing}) versus drawbacks (\emph{negative framing}) can affect perceptions and decisions \parencite{Kuehberger1998,Cacciatore2015,Flusberg2024}.
However, there is a gap if and to what degree framing shapes the evaluation of collaborative robots (cobots) in industrial settings.

This study contributes to this challenge of social robotics by studying how framing shape the perception of robots.
Specifically, it investigates how message framing (positive vs. negative) affects the public perception of HRC and the relationship between perceived risks, perceived benefits, and the attributed value of HRC using a sample representative for the German working population.
By exploring how individuals integrate different framing cues, we contribute to a more nuanced understanding of acceptance in the age of digitalized, human-centered production and can better anticipate reactions and design communication measures if HRC is implemented.

\section{Related Work}
\label{section:related}

The Czech author Karel Čapek coined the term \enquote{\emph{robot}} around 1920, meaning \enquote{\emph{forced worker}} \parencite{etymonline_robot}.
More formally, the International Organization for Standardization (ISO) defines a robot as \enquote{\ldots automatically controlled, reprogrammable, multipurpose, manipulator programmable in three or more axes, which may be either fixed in place or mobile for use in industrial automation applications} \parencite{ISO8373}.
A broader definition describes a robot as a \enquote{constructed system that displays both physical and mental agency, but is not alive in the biological sense} \textcite{Richards2016}.

While robots have been part of science fiction for decades, such as \emph{R2D2} from Star Wars, \emph{Terminator}, or \emph{Marvin} from The Hitchhiker's Guide to the Galaxy, they have long left that domain. Industrial use began with the \emph{Unimate} in a General Motors factory in 1961 to move hot pieces after die casting \parencite{Unimate1961}.
While they are assumed to be safe, severe accidents do happen:
The first fatal accident happened in 1979 at the Ford Motor Company \parencite{RobotAccident1979}.
Between 2015 and 2024, 77 robot-related accidents were reported: 54 from stationary robots with injuries such as amputations and fractures to head and torso and 23 from mobile robots with mainly leg and foot fractures \parencite{Sanders2024}.
Hence, risk perceptions are a concern and need to be considered.


In industrial production, the integration of robotics into workflows has brought a new paradigm towards \enquote{cobots}, where humans and robots interact in shared spaces and perform interdependent tasks \parencite{Michalos2018,Matheson2019}.
As the degree of interaction increases, human–robot relations are differentiated into coexistence, cooperation, and collaboration. In contrast to coexistence and cooperation, human–robot collaboration (HRC) inherently requires physical contact between humans and robots \parencite{Schmidtler2015}).
In the later, the collaboration typically involves shared workspaces and tasks, requiring advanced sensors and intelligent systems to ensure safe and effective interaction \parencite{Vysocky2016}.
This form of HRC is central to Industry 4.0 as it enables greater flexibility and responsiveness on the shop floor \parencite{Villani2018,Bauer2023}.
While the technical capabilities of HRC systems have progressed rapidly, studies suggest that successful deployment depends on worker acceptance and trust \parencite{Coronado2022,Hoff2015,Liehner2021,Broehl2019}.
Much of the foundational research in technology acceptance stems from models like the Technology Acceptance Model (TAM) \parencite{Davis1989} and the Unified Theory of Acceptance and Use of Technology (UTAUT) \parencite{Venkatesh2003}. These models emphasize perceived usefulness and ease of use as drivers of adoption.
However, in the context of robotics, especially in safety-critical environments, affective and risk-related factors, such as trust and perceived danger, play a more dominant role \parencite{Hancock2011,Sheridan2016,Broehl2019}.
To this end, trust, perceived risks (or safety respectively), and perceived benefits emerged as key predictors of positive user attitudes in industrial settings, especially when people collaborate in close proximity with robots \parencite{Maurtua2017,Murashov2016,Broehl2019}.
 Similarly, broader research on robot and AI perception highlights that transparency, education, and governance are crucial in shaping trust and mitigating perceived risks, reinforcing the importance of these factors for human-robot interaction in Industry 4.0 contexts \parencite{Hilliard2024}.

\subsection{Framing}

Media theory argues that media influences beliefs, attitudes, and behaviors by different (and not always clearly distinguishable) mechanisms \parencite{Hoewe2020,Scheufele1999,Scheufele2000}:
Media outlets make selected topics more salient than others through \emph{agenda setting} and thereby influence which topics people consider to be more important than others (\emph{which} topics are presented).
\emph{Media framing} describes in which frame a certain aspect is presented.
Selected contextual information on a topic can shift the individuals' attention to different aspects (in which \emph{context} or frame topics are presented).
\emph{Media priming} describes \enquote{the effects of the content of the media on people’s later behavior or judgments} \parencite[p.~75]{RoskosEwoldsen2009} (\emph{how} topics are presented).

Media framing and priming involve selecting certain aspects of reality and making them more salient in communicative contexts, thereby shaping how audiences interpret and evaluate information \parencite{Entman1993,Cacciatore2015}.
Especially in contexts involving risk or uncertainty, like with emerging technologies, media frames shape attitudes and behaviors \parencite{Scheufele2006,Chong2007}.
These frames can be studied as either independent or dependent variables \parencite{Scheufele1999}.
The former examines the influence of media frames on the formation of attitudes, beliefs, and behavior (as independent variable).
The latter analyses how media frames are constructed and  implicitly influenced by the socio-cultural background of their creators (as dependent variable).
Although it certainly makes sense to also address the latter in the context of Responsible Research and Innovation (i.e., to understand the role of the researchers on the design process; see \textcite{Dwyer2023} for an analysis of how media frames on AI and robotics changed during the Covid-19 pandemic), this work is concerned with media frames as an independent variable (i.e., how the frame changes perception).


A meta analysis showed that media priming influences judgments or behaviors across many domains \parencite{RoskosEwoldsen2007} as the following examples outline:
The way health messages are framed (as highlighting benefits or emphasizing risks) can shape people’s decisions, but their effectiveness depends on whether the behavior feels risky and how individuals interpret the message \parencite{Rothman1997}.
Positively framed messages are modestly more effective than negatively framed ones in promoting prevention behaviors like skin cancer protection and physical activity, though they show little advantage when it comes to shaping attitudes, intentions, or detection-related actions \parencite{Gallagher2011}.
Framing climate change messages in terms of gains rather than losses can foster more positive attitudes and heighten perceived severity of the issue, though these effects may be limited by lower fear responses and memory for positively framed content; additionally, emphasizing distant and social impacts can further strengthen support for climate action \parencite{Spence2010}.
Framing can influence participation in scientific studies and lead to biased samples \parencite{Tal2018}:
If the call for participation in a survey is framed as \enquote{contribution to research}, participation was lower than if it was framed about \enquote{comparison with others}.
While gender diversity in leadership roles was generally deemed important, framing the gender gap as the overrepresentation of men increased perceptions of injustice compared to framing it as the underrepresentation of women \parencite{Liaquat2025}.
In the domain of human–technology interaction, framing can affect general evaluations and the salience of specific attributes such as perceived risks or benefits \parencite{Liu2009,Binder2010}.
The portrayal of automated vehicles (AVs) in media demonstrates that impact.
News stories about AV accidents can disproportionately affect public perception, making AVs seem less safe despite their actual safety records \parencite{Wong2019}.

The framing of robots also influence how humans interact with them.
For instance, presenting a robot as a social agent versus a machine-like entity affects children's social behaviors, such as conversation, mimicry, and prosocial actions \parencite{KoryWestlund2016}.
A study suggests that multiple communication roles of a robot realized through framing can alter public behavior towards the robots.
Different depending on the presented communication frame of a robot different human interaction rituals are replicated \parencite{Fortunati2018}.
\parencite{Biermann2021} studied if trust in and perception of robots depends on their framing using visual appearance and usage context.
Specifically, they showed that a) the 'intention to use' as a measure for social acceptance was higher for production robots, b) both context (production or care context) and appearance (functional vs.~anthropomorphic) influenced trust and that trust was highest for production robots with a functional appearance, and c) differences in attributions between functional and anthropomorphic robots were much bigger in the care context.

While media framing has been shown to influence public attitudes toward technologies like Artificial Intelligence (AI) and automation \parencite{Bingaman2021,Ho2024}, studies in the domain of HRC and interaction are scarce.
However, HRC is a building block of future smart factories, where human workers and robots work and collaborate in close spatial proximity to perform production tasks \parencite{Villani2018,Coronado2022}.
As balancing industrial productivity gains with human values for the workers is crucial, we want to contribute to an understanding how social valuation of HRC is shaped. In this work, we address the scientific gap if and to what degree media framing influences value formation of HRC. Results should contribute to an understanding how media framing helps designing effective communication strategies that balance risk and benefit information for a successful integration of HRC in the digital transformation of production.

\subsection{Research Questions and Hypotheses}
\label{section:hypotheses}

To evaluate the influence of media framing on the perception of HRC, we developed a survey with an experimental component.
The following hypotheses guide our research.

\begin{itemize}
\item[H1:] We expect a significant and strong main effect of media framing (negative vs. positive framing) on the evaluation of the framing stimuli in form of a newspaper article (H1, as a manipulation check if the framing got actually noticed). 
\item[H2:] We postulate a weak, but significant effect of media priming on the overall perception of human-robot collaboration (H2).
\item[H3:] We postulate a weak effect of media framing on the risk-benefit tradeoff, meaning that media framing change how individuals integrate perceived risks and benefits of HRC in the overall formation of value judgements.
\end{itemize}

%
%
\section{Method}
\label{section:method}

\begin{figure}[bh]
	\centering
	\includegraphics[width=\textwidth]{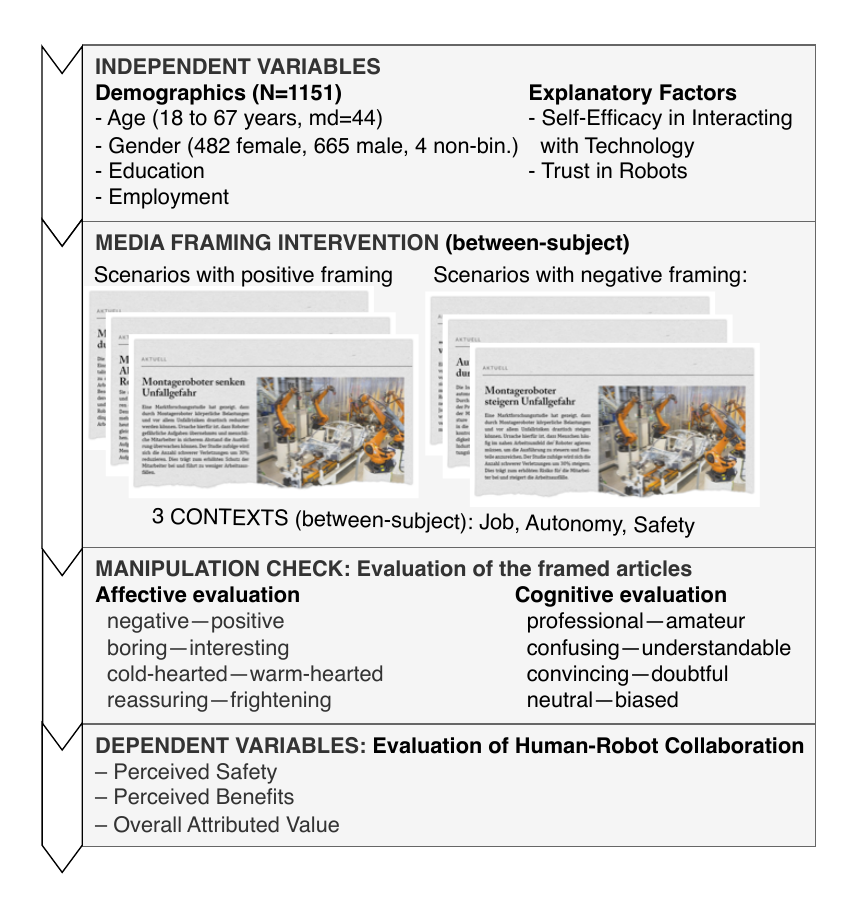}
	\caption{Overview of the design of the online survey, including demographic and personality measures, exposure to framed news articles (between-subjects design), a framing manipulation check, and evaluation of human–robot collaboration.}
	\label{figure:surveydesign}
\end{figure}

To investigate the hypothesized effects of media framing, we designed an online survey with four main parts that will be detailed in the following sections (illustrated in Figure \ref{figure:surveydesign}):
First, we surveyed the participants' demographics and selected personality traits.
Second, we presented one out of six mock-up newspaper articles on cobots in a production environment.
The articles had a) positive or negative framing and b) addressed one of the three different contexts autonomy, safety, and employment security ($2 \times 3$ between-subject design).
Third, to check if framing is noticed by the participants, we asked them to evaluate the articles on both affective and cognitive aspects (as a manipulation check).
Fourths, we asked participants to assess HRC in general, in terms of perceived risks, benefit, and overall value.

To increase transparency and reproducibility of our study, we pre-registered our methodology, research hypotheses, and desired sample with AsPredicted (\url{https://aspredicted.org/j336-r458.pdf}).
It is part of a larger survey on the perception of HRC and shared participants and independent variables with two other studies, as documented in the preregistration.
All data, materials, and analysis scripts are publicly available via an Open Science Foundation repository (BLINDED).
At the start of the survey, participants were informed about the nature of the study and data usage and then provided informed consent.

\clearpage

\subsection{Demographics and Individual Difference Factors}
\label{section:method:explanatory}

As explanatory user factors we queried the participants' \emph{age in years}, \emph{gender}, and their highest educational attainment.
Gender was measured as either \emph{male}, \emph{female}, \emph{non-binary}, and \emph{prefer not to say} (based on an operationalization suggested by \parencite{Spiel2019}).
For querying the highest educational attainment, we used \emph{no formal education}, \emph{vocational training}, \emph{master craftsman diploma}. \emph{university degree} and \emph{PhD} as options.
Employment status was assessed by asking participants to indicate their current situation from the following options: \emph{employed}, \emph{self-employed}, \emph{unemployed/seeking work}, \emph{student}, \emph{retired}, \emph{homemaker}, or \emph{other}.
Experience with robots was assessed by asking whether participants encounter robots in their work (\enquote{Do you interact with robots in your job?} with the binary response options \emph{yes} or \emph{no}.

\emph{Self-Efficacy in interacting with technology (SET)} describes the individual attitude towards cognitively demanding tasks and how individuals like to pursue them \parencite{Beier1999}.
Based on the original scale, we created a shortened version by selecting the four items with the highest item-total-correlation.
An example item is \enquote{I can solve many of the technical problems I encounter on my own} and the participants responded on six levels of agreement ranging from \emph{strongly disagree} to \emph{strongly agree}.
This scale had high internal reliability (Cronbach's $\alpha=.83$).

We further queried \emph{trust in robots} (TIR) using a five item scale based on the trust in automation scale by \textcite{Jian2000}.
This scale achieved very high internal reliability (Cronbach's $\alpha=.82$).

\subsection{Independent Variable: Media Framing}
\label{section:method:independent}

\begin{figure}[bth]
	\centering
	\includegraphics[width=\textwidth]{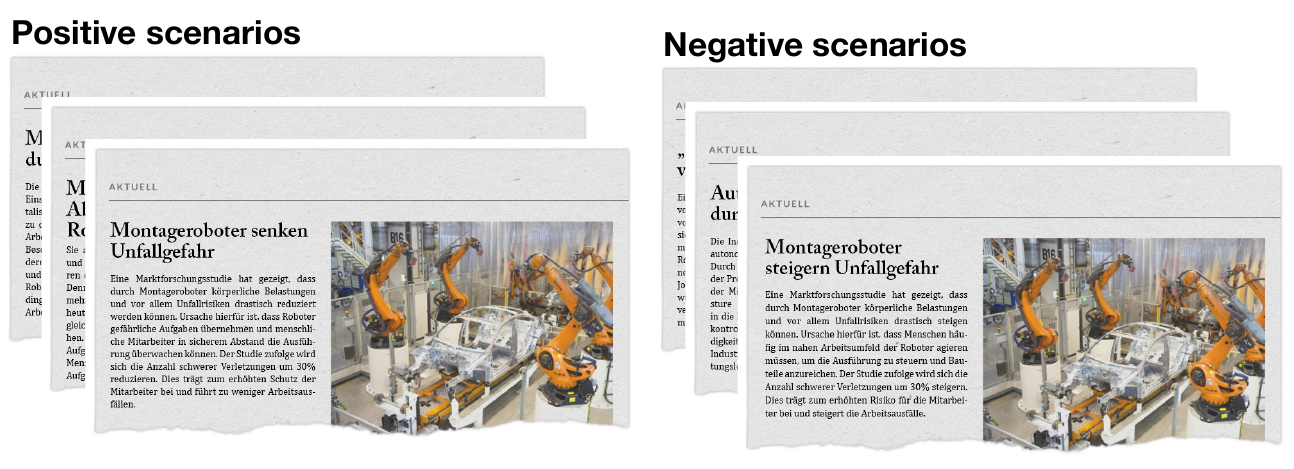}
	\caption{Three positive and three negative (autonomy, safety, job security) news paper articles used as stimuli for media priming. Table \ref{tab:framing-scenarios} shows the texts of each scenario.}
	\label{figure:directscenarioevaluation}
\end{figure}

Following \textcite{Nan2018}, we used positively and negatively framed newspaper articles to examine how media framing influences perceptions of robots.
To enhance the robustness of our study, the articles addressed three common contexts associated with human–robot collaboration: autonomy (gain/loss) \parencite{vanDijk2023,Glawe2025}, safety (gain/loss) \parencite{Rubagotti2022,Broehl2019}, and job security (gain/loss) \parencite{Nedelkoska2018}.
Hence, participants were randomly assigned to read one of six newspaper articles discussing the effects of cobots on workers in a $2\times3$ between-subject design.
We designed the articles with the positive and negative framing to be complementary.
Each scenario was presented in form of a newspaper article with a headline, three to five descriptive sentences, and the same image for all scenarios (see Figure \ref{figure:directscenarioevaluation}).
The assignment of the participants to the three contexts (autonomy/safety/job security) and two frame polarizations (positive/negative) was random.

Note that our operationalization rather aligns with media theory \parencite{Scheufele2006,Cacciatore2015} than with prospect theory \parencite{Gilovich2002,Kahneman1979} \footnote{The later involves presenting \emph{identical} information (\emph{equivalence framing}) in different ways (e.g., \emph{the treatment saves 40 of 50 lives} vs. \emph{10 of 50 people will die}), whereas media framing theory involves highlighting issue in an distinct manner (\emph{emphasis framing}, example from this study: \emph{the number of serious injuries will decrease by 30\%} vs. \emph{the number of serious injuries will increase by 30\%}).}.
Table \ref{tab:framing-scenarios} shows the text of each presented scenario.

\subsection{Dependent variables: Measuring the Framing Effect}\label{section:method:dependent}

We measured the effect of media framing by asking the participants to evaluate 1) the framing stimuli and then 2) by evaluating their perception of HRC in general.

First,
we asked the participants to evaluate the news paper articles to check if the presented stimuli were actually perceived differently (manipulation check).
Hereto, we used 8 six-point semantic differential items capturing both \emph{affective} (i.e., \enquote{negative --- positive}, \enquote{boring --- interesting}, \enquote{cold-hearted --- warm-hearted}, \enquote{reassuring --- frightening}) and \emph{cognitive} aspects (i.e., \enquote{professional --- amateur}, \enquote{confusing --- understandable}, \enquote{convincing --- doubtful}, \enquote{neutral --- biased}) of media perception.

Second,
afterwards, we queried the participants general attitudes towards HRC using three self-developed scales.
Our research focus lies in the impact of media framing itself, rather than the detailed assessment of cobot perception.
Therefore, instead of using a comprehensive but longer acceptance scales (such as the established, validated, yet long instrument by \textcite{Broehl2019}), we developed three short scales for the constructs \emph{perceived risks}, \emph{perceived benefits}, and \emph{attributed value}, each measuring the construct with three items.
\emph{Perceived risks} (as the inverse of perceived safety) captures individuals’ emotional sense of risks during interactions, which is essential for trust and willingness to engage with collaborative robots in shared spaces \parencite{Rubagotti2022}.
\emph{Perceived benefits} reflect the expected advantages of using cobots and are known to influence acceptance and positive attitudes \parencite{Broehl2019}.
\emph{Attributed value} represents the perceived trade-off between benefits and risks, and serves as the target variable in value-based technology acceptance models \parencite{Kim2007}.

Each scale includes positively and negatively worded items (see Table \ref{tab:robotevaluations} in the Appendix).
Example items include \enquote{Interacting with collaborative robots poses significant risks} for perceived risks, \enquote{Collaborative robots increase efficiency and productivity} for perceived benefits, and \enquote{Collaborative robots are a positive addition to many workplaces} for attributed value.
The three short scales achieved high internal reliability (Cronbach's $\alpha$ for perceived risks: .72, perceived benefits .72, and overall perceived value .78).
Consequently, despite their brevity, the scales enable reliable measurements.
The items and scale characteristics are detailed in Table \ref{tab:robotevaluations} in the Appendix.

Since the study involved paid participants who may have limited motivation to  contributing to scientific progress, we included three items as reactive attention checks (e.g., \enquote{Attention check: Please select \enquote{Agree}}) at different positions in the survey.
The order of the items was randomized within the respective blocks of the survey.

\subsection{Statistical Procedures}
\label{statistical-procedures}

We analysed the data using parametric and non-parametric methods depending on the measurement model of the variables and report arithmetic means ($M$), standard deviations $\text{SD}$, and the 95\% confidence interval (CI-95).
We use Pearson's $r$ (parametric), Spearman's $\rho$ (non-parametric), and Kendall's $\tau$ correlation coefficient, multiple linear regressions, and (Multivariate) Analysis of Variance ((M)ANOVA) using Pillai’s trace ($V$) as the omnibus statistic.
Missing data were handled using test-wise deletion. All statistical analyses were conducted in R version 4.5.1.
Following \textcite{Cohen1988} effect sizes are reported as $\eta^2$ and interpreted as small ($\eta^2<.04$), medium ($\eta^2<.14$), or large ($\eta^2>.14$) 
The Type 1 error rate (level of significance) was set at $\alpha= 0.05$, a conventional threshold in social science research to balance the risk of Type I error while maintaining sufficient sensitivity to detect meaningful effects \parencite{Cohen1994,APA2020}.
The authors used a Large Language Model (ChatGPT and Gemini) for editorial refinement and assistance in coding the analysis.
All outputs were reviewed, validated, and revised by the authors.
The materials, data, and notebook with our analysis is available on OSF: \url{https://osf.io/bd6xm/overview?view_only=a9f8436a45d842a486fb1fe2a0422e50}.
The research plan was pre-registered with AsPredicted: \url{https://aspredicted.org/j336-r458.pdf}.

%
We conducted a priori power analyses with G*Power 3.1 to determine the minimum sample size required for a) the manipulation check and b) the effect on the risk, benefit and value perception, and c) the linear regression analysis of the risk-benefit tradeoff.
The power analyses were designed to detect small ($f^2 = 0.05$), medium ($f^2=0.15$), and large effect sizes ($f^2 = 0.35$) at a significance level of $\alpha = .05$ (5\%) and a desired statistical power of $0.80$ \parencite{Cohen1988}.
For a) we expected a large effect size of $f^2(V)=0.35$ or greater and the power analysis for a MANOVA with 2 groups and 8 dependent variables yielded a minimum sample size of $n=52$ participants.
%
For b) we expected a medium effect size of $f^2(V)=0.15$ or greater and the power analysis for a MANOVA with 2 groups and 3 dependent variables yielded a minimum number of $n=111$ participants.
For c) we expected at least a small effect size ($f^2=0.05$) and the power analysis for a linear regression with two predictors and the increase of $R^2$ through a third yielded a minimum sample size of $n=159$ participants.

\subsection{Description of the Sample} \label{section:sample}

We recruited participants representative of the German working population in terms of age and gender via an online panel.
Participants received an incentive of 0.75 to 1.25 EUR for completing the survey.
Although this amount is below the standard hourly wage, it was calibrated to acknowledge participants’ effort without creating undue inducement that could compromise voluntary consent.
After filtering the data for straightlining, speeders (faster that 50\% of the median completion time of 16.8 min.), and incorrectly responding to the attention items \parencite{Leiner2019}, the final sample consisted of 1151 of the originally 1507 started surveys (the dropout rate of 23.6\% is comparable to similar studies, unfiltered data and filtering code available on OSF).

Of the 1151 participants, 482 identified themselves as female (41\%), 665 as male (58\%), four (\textless1\%) as non-binary, and zero preferred not disclose their gender identity.
Ages ranged from 18 to 67 years, with a median age of 44 years (mean age 44.1 years, SD = 13.1 years).

Most participants were employed (n = 967), while smaller groups identified as pupils or students (n = 62), retired (n = 50), unemployed or seeking work (n = 27), self-employed (n = 22), homemakers (n = 10), or reported \emph{other} as employment status (n = 13).
This distribution reflects a sample primarily composed of working individuals, with representation from other life and career stages.
The sample included participants with a range of educational backgrounds.
The majority held a vocational training qualification (n = 635), followed by those with a university degree (n = 303). A smaller portion reported having a master craftsman qualification (n = 117), no formal training yet (n = 71), or a doctorate (n = 25).
This reflects a broad cross-section of educational attainment levels in the German population.

Age was not significantly correlated with self-efficacy in interacting with technology ($p =.860$) or trust in automation ($p=.948$). 
Gender, was significantly associated with both self-efficacy ($\tau = .225$, $p < .001$) and trust in robots ($\tau=.161$, $p<.001$), with women reporting lower self-efficacy and lower trust in robots compared to men.
Self-efficacy in interacting with technology was strongly correlated with trust in automation ($r = .488$, $p < .001$), suggesting that individuals who reported higher trust also reported higher self-efficacy and vice versa.
These findings highlight the role of gender differences in attitudes toward technology and the strong interplay between trust and self-efficacy in automation contexts.

578 (50\%) participants received an article with a negative frame and 573 (50\%) an article with a positive one.
By design, the randomly drawn framing condition is unrelated to the other independent variables ($p>.05$).

\section{Results}\label{section:results}

The results section is structured as follows.
First, we analyse if the participant reacted to the framing stimuli by checking if the presented news articles are perceived differently.
Second, we check if media framing has an overall effect on perceived risks, perceived benefits, and the overall valuation of cobots in production.
Lastly, we analyse if the risk-benefit tradeoffs are affected by media framing.

\subsection{Did Media Framing Work? --- Evaluation of the Articles used as Stimuli}
\label{section:results:framing}

To check if media framing was noticed by the participants, we checked for a significant multivariate response based on the administered framing.
We calculated a Multivariate Analysis of Variance (MANOVA) with the polarization of the articles (positive or negative frame) and their context (autonomy, safety, job security) as two independent factors and the 8 article evaluation items as dependent variables.
The overall effect of the article's framing on the evaluation is strong and significant ($V=.383$, $F(8,944)=73.189$, $p<.001$).
While the specific context of the framing had no significant overall effect on the evaluation ($V=0.019$, $F(16,1890)=1.130$, $p=.321$), there is a weak but significant interaction between framing and scenario ($V=.044$, $F(8,944)=5.463$, $p<.001$). 

\begin{figure}
	\centering
	\includegraphics[width=\textwidth]{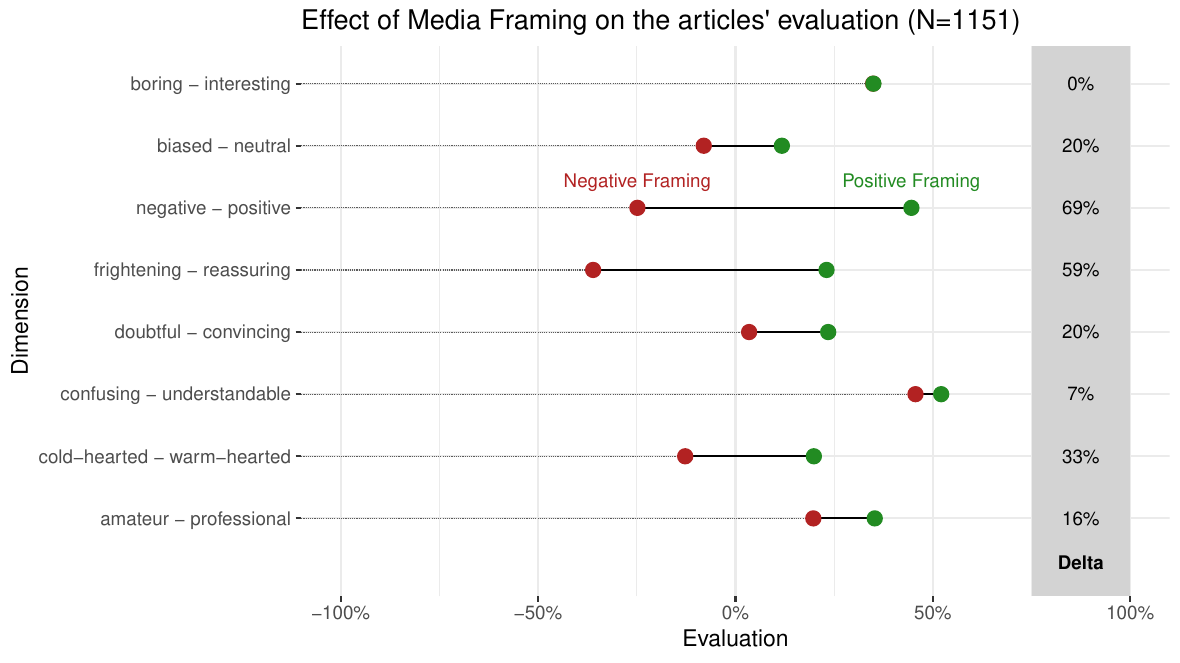}
	\caption{Effect of media framing on the evaluation of the newspaper articles: Articles with a positive frame are rated significantly different on most dimensions compared to articles with a negative frame.}
	\label{fig:dumbellframing}
\end{figure}

Figure \ref{fig:dumbellframing} illustrates the effect of the article's framing on the evaluation of the article.
As the figure suggests, all of the articles' evaluation dimensions but the dimension \enquote{boring---interesting} were significantly affected by the articles' framing (almost all at $p<.001$, only \enquote{confusing---understandable} at $p<.01$).
The greatest differences emerged for the affective evaluation of the newspaper article, specifically for the items pairs \enquote{positive---negative} (69\% difference), \enquote{frightening---reassuring} (59\% difference), and \enquote{cold-hearted---warm-hearted} (33\% difference).
As expected, the positive framing yielded a more positive cognitive and especially affective evaluation of the newspaper articles compared to the negative framing.
Therefore, we conclude that the framing intervention is effective in shaping the participants' views.

\subsection{Effect of Media Framing on Absolute Risk-Benefit Perception and Value Attribution of Cobots}

Next, we analyse the absolute effect of media framing on the perception of risk and benefit, as well as  value attribution.
We calculated a MANOVA with the frame type (positive/negative) and context as independent variables and perceived risks, benefits, and overall attributed value as dependent variables.
The MANOVA yielded a weak but significant overall effect of framing on the three dependent variables ($V=.050$, $F(3, 949)=16.577$, $p<.001$).
The main effect Framing is significant ($V=.050$, $F(3, 949)=16.5$, $p<.001$) but the main effect scenario is not  ($V=.013$, $F(6,1900)=1.999$, $p=0.063$).
Hence, framing (positive, negative) does alter the overall absolute levels of perceived risks, benefits and value of industrial HRC, while the shown scenarios (autonomy, safety, employment) do not.

The interaction of framing and scenario on the three dependent variables is weak but significant ($V=.044$, $F(3,949)=14.541$, $p<.001$).
In particular, the framing condition had a significant effect on the perceived risks ($F(1,951)=38.4$, $p<.001$), the perceived benefits ($F(1,951)=5.9$, $p=.015$), and overall value of HRC ($F(1,951)=10.6$, $p=.001$).
The scenario only had a significant effect on the perceived risks ($F(1,951)=3.029$, $p=.049$), but not on benefit perceptions ($F(2,951)=0.305$, $p=.737$) or value attributions ($F(2,951)=0.965$, $p=.381$).
Lastly and likewise, the interaction between framing and scenario is only significant for the perceived risks ($F(1,951)=17.01$, $p<.001$), but not for perceived benefits ($F(1,951)=0.018$, $p=.894$) or attributed value ($F(1,951)=0.110$, $p=.739$).  
Table \ref{tab:robotevaluations} in the Appendix lists the 9 items and their characteristics.

\begin{figure}
	\includegraphics[width=\textwidth]{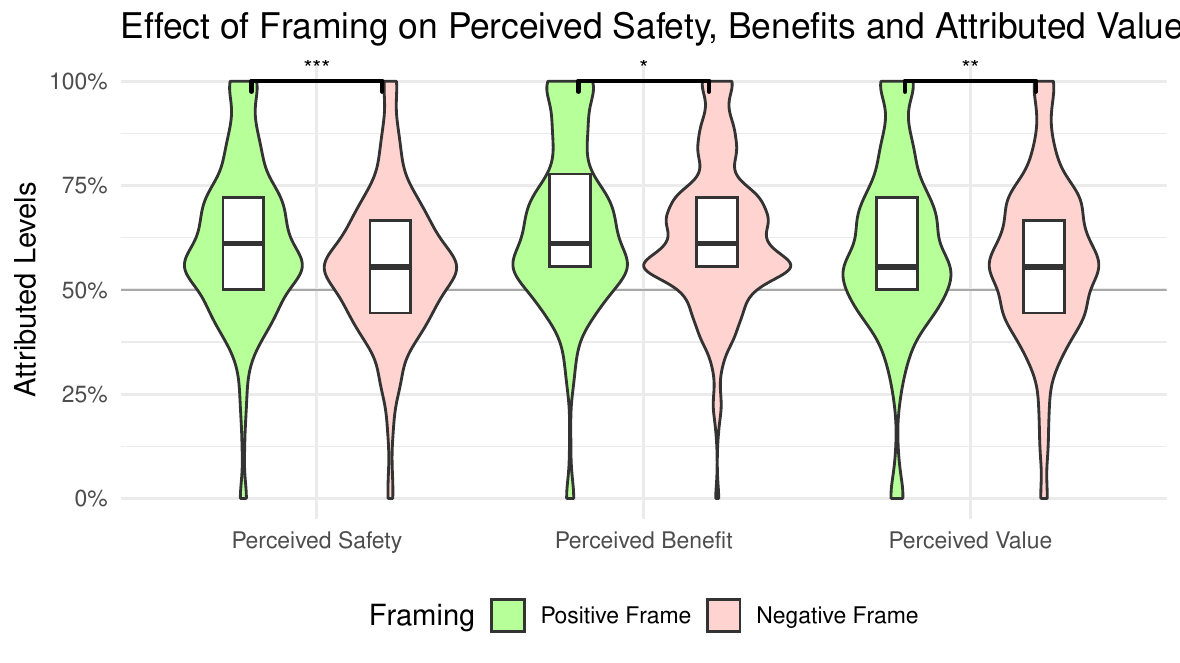}
	\caption{Overall significant effect of media framing on the absolute evaluations of perceived risks, benefits, and attributed value of HRC. The violin plot illustrates the distribution of the responses; the lines in the plot illustrate the 25\%, 50\% (median), and 75\% percentile of the distribution.}
	\label{fig:boxplotTAMafterFraming}
\end{figure}

In line with our expectations, perceived risks were rated higher in the negative framing condition (M=55.8\%, SD=17.2\%) than with positive framing (M=61.4\%, SD=18.8\%).
Likewise, the perceived benefits were rated higher with positive (M=64.8\%, SD=19.7\%) rather than negative framing (M=62.8\%, SD=17.3\%).
Lastly, attributed value was higher in the positive frame (M=60.0\%, SD=20.2\%) than in the negative one (M=56.8\%, SD=19.2\%).
Figure \ref{fig:boxplotTAMafterFraming} illustrates this effect.
 
\subsection{Analysis of the Risk-Benefit Tradeoff}

Next, we analyse if framing influences how individuals' integrate perceived risk and benefit for forming their overall value judgements.
For that, we calculated a multiple linear regression with attributed value as dependent variable and perceived benefits and as metric predictors and the framing condition as a binary predictor.
If framing has a significant influence, the regression models for both positive and negative framing are significantly different and can be analysed independently (cf. \textcite{Chow1960}).

The significant regression model explains 66\% of the variance of attributed value ($F(7,1143)=311.4$, $p<.001$, $r^2=.656$).
While there is no significant main effect of framing on the overall attributed value ($\beta=0.082$, $p=.143$), there is a significant three-way interaction between perceived risks, perceived benefits, and the framing condition ($\beta=0.330$, $p=.023$).
This indicates that the tradeoff between perceived risks and benefits for forming value judgements is significantly affected by the framing condition, even though framing does not directly affect the influence of perceived risks and benefits on value attribution.
Table \ref{tab:regressioncombined} in the appendix presents the details of this regression analysis.

Based on the significant effect of framing on the tradeoffs, we assume significantly different regression models for both positive and the negative framing that will be analysed individually in the following.

\begin{figure}[bt]
	\centering
	\includegraphics[width=\textwidth]{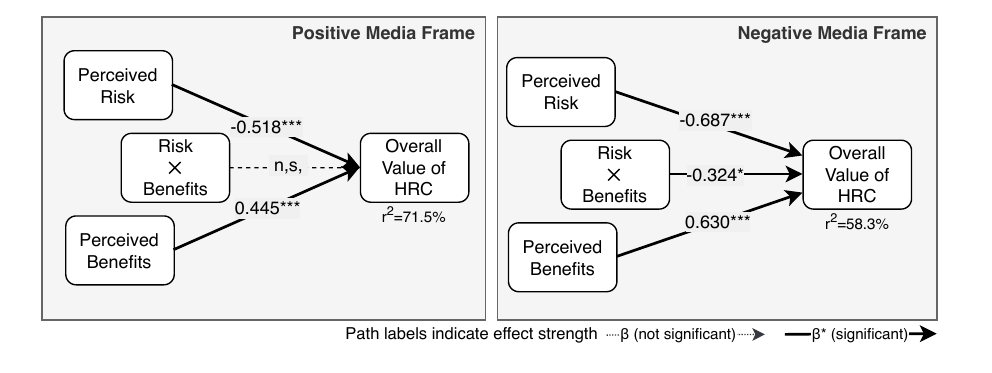}
	\caption{Significant regression coefficients of perceived risks, perceived benefits, and their interaction on overall attributed value for the positive (left) and negative (right) framing conditions.}
	\label{figure:mainregression}
\end{figure}

\emph{In the negative framing condition}, we conducted a multiple regression analysis to examine the effects of perceived risks, perceived benefits, and their interaction on the overall attributed value of HRC.
The model was statistically significant ($F(3, 574)=268$, $p < .001$), explaining approximately 58.3\% of the variance in attributed value ($R^2 = .583$).
Both perceived risks ($\beta = -0.69$, $p < .001$) and perceived benefits ($\beta = 0.63$, $p < .001$) were significant predictors of attributed value.
Importantly, a significant negative interaction between risks and benefits ($\beta = -0.32$, $p = .009$) indicated that the combined influence of high perceived risks and high perceived benefits was less than the sum of its parts.
This suggests a potential diminishing returns effect, wherein the positive impact of one factor is attenuated when the other is already high.
Table \ref{tab:regression:lossframe} and Figure \ref{figure:mainregression} (right) show the full regression model for the negative framing condition.

\begin{table}[ht]
\centering
\begin{threeparttable}
\caption{Multiple Linear Regression Results for Predicting Overall Value based on Perceived Risks and Perceived Benefits in the \emph{Negative frame} condition.\label{tab:regression:lossframe}}
\begin{tabular}{lrrrc}
\toprule
\textbf{Variable} & \textbf{$\beta$} & \textbf{SE} & \textbf{T} & \textbf{$p$} \\
\midrule
(Intercept)        & -0.091 & 0.047 & -1.949 & .052 \\
Perceived Risks           & -0.687  & 0.089 & 7.683  & $<.001^{***}$ \\
Perceived Benefit          & 0.630  & 0.080 & 7.883  & $<.001^{***}$ \\
Risks $\times$ Benefit & -0.324 & 0.123 & -2.629 & $.009^{**}$ \\
\midrule
\multicolumn{5}{l}{$F(3, 574)=268$, $p < .001$, $R^2=.583$ ($R_{\text{adj}}^2=.581$)} \\
\bottomrule
\end{tabular}
\begin{tablenotes}
\small
\item \textit{Note:} $^{***}p<0.001$, $^{**}p<0.01$
\end{tablenotes}
\end{threeparttable}
\end{table}

\emph{In the positively framed condition}, the multiple regression analysis revealed a statistically significant model ($F(3, 569) = 475.1$, $p < .001$, accounting for 71.5\% of the variance in the overall attributed value of HRC ($R^2 = .715$).
Both perceived risks ($\beta = -0.52$, $p < .001$) and perceived benefits ($\beta = 0.45$, $p < .001$) were significant positive predictors of attributed value.
In contrast to the negatively framed condition, the interaction between risks and benefits was not significant ($\beta = 0.006$, $p = .941$), indicating that their effects were additive but independent in this framing context. 
The intercept was statistically zero, suggesting no inherent bias in attributed value when both predictors were at their neutral levels.
Table \ref{tab:regression:gainframe} and Figure \ref{figure:mainregression} (left) show the full regression model for the positive framing condition.

\begin{table}[ht]
\centering
\begin{threeparttable}
\caption{Multiple Linear Regression Results for Predicting Overall Value based on Perceived Risks and Perceived Benefits in the \emph{positive frame} condition.\label{tab:regression:gainframe}}
\begin{tabular}{lrrrc}
\toprule
\textbf{Variable} & \textbf{$\beta$} & \textbf{SE} & \textbf{T} & \textbf{$p$} \\
\midrule
(Intercept)        & -0.009  & 0.032 & -0.278 & .781 \\
Perceived Risks           & -0.518   & 0.064 & 8.131  & $<.001^{***}$ \\
Perceived Benefits          & 0.445   & 0.060 & 7.388  & $<.001^{***}$ \\
Risks $\times$ Benefits & 0.006   & 0.080 & 0.075  & .941 \\
\midrule
\multicolumn{5}{l}{$F(3, 569)=475$, $p < .001$, $R^2=.715$ ($R_{\text{adj}}^2=.713$)} \\
\bottomrule
\end{tabular}
\begin{tablenotes}
\small
\item \textit{Note:} $^{***}p<0.001$
\end{tablenotes}
\end{threeparttable}
\end{table}

\subsection{Exploratory Analysis of the Influence of User Diversity}

In the following, we analyse how individual-level predictors (meaning individual differences or factors of user diversity) interact with message framing in shaping overall evaluations of HRC.
As this part of the analysis was not preregistered, we did not calculate a full confirmatory path model but instead performed an exploratory hierarchical multivariate multiple regression.
The dependent variables were perceived benefits and perceived risks, with message framing (positive vs. negative) treated as a nesting factor and thus effectively running two parallel regressions for each frame condition.
In the first step, we include age and gender as demographic predictors of perceived risks and benefits.
In the second step, we added the two additional personality variables: self-efficacy in interacting with technology and trust in robots.

In the first model (based on demographics), the omnibus test revealed a significant effect of framing on the dependent variables perceived risks and benefits ($V = 0.047$, $F(2, 945) = 23.75$, $p < .001$).
However, the interactions between framing and age ($V = 0.061$, $F(4, 1892) = 2.26$, $p = .061$) and framing and gender ($V = 0.008$, $F(4, 1892) = 1.93$, $p = .103$) were not statistically significant.
This suggests that message framing significantly influences perceptions of HRC, regardless of age and gender.

In the second model (with personality variables), the omnibus test, again, showed a significant main effect of framing ($V = 0.061$, $F(2, 941) = 30.65$, $p < .001$).
Additionally, the interactions of framing with age ($V = 0.011$, $F(4, 1884) = 2.63$, $p = .033$), gender ($V = 0.011$, $F(4, 1884) = 2.55$, $p = .038$), self-efficacy in interacting with technology ($V = 0.129$, $F(4, 1884) = 32.36$, $p < .001$), and trust in robots ($V = 0.257$, $F(4, 1884) = 69.58$, $p < .001$) were all statistically significant.
Further, the comparison of the two models
showed a strong and significant improvement in model fit through adding the two personality predictors ($V = 0.331$, $F(8, 1884) = 46.69$, $p < .001$).
This indicates that the inclusion of self-efficacy and trust in robots substantially and significantly increases the explained variance in perceptions of risks and benefits in HRC.

\begin{figure}[bt]
	\centering
	\includegraphics[width=\textwidth]{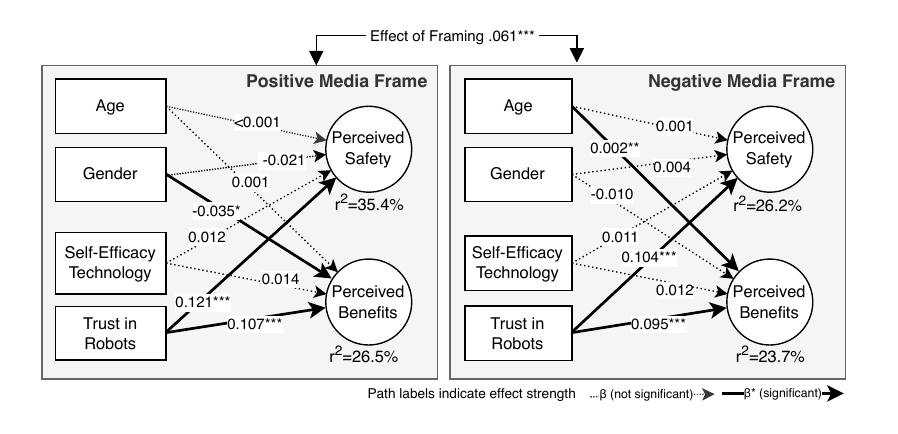}
	\caption{Illustration of the significant exploratory Multivariate Multiple Linear Regression results for demographics and explanatory variables on Perceived Risks and Value for both positive (left) and negative framing (right).}
	\label{figure:userdiversityregressionresults}
\end{figure}

In the positive frames, the analysis of the contributions of the predictors suggests that perceived risks were significantly influenced by trust in robots ($\beta = -0.121$, $p < .001$), while all other predictors had no significant effect.
Perceived benefits were significantly affected by gender ($\beta = 0.035$, $p = .023$), with women seeing fewer benefits than men, and by trust in robots ($\beta = 0.108$, $p < .001$), with more robot trusting persons seeing more benefits.
Figure~\ref{figure:userdiversityregressionresults} (left) visualizes the results.

In the negative frames, trust in robots was the only significant predictor of perceived risks ($\beta = -0.104$, $p < .001$), with people having higher trust in robots reporting lower perceived risks.
Both age ($\beta = 0.002$, $p = .008$) and trust in robots ($\beta = 0.095$, $p < .001$) significantly increased the perceived benefits, with older (!) and more robot-trusting people seeing more benefits in colaborative robots.
Figure~\ref{figure:userdiversityregressionresults} (right) illustrates these results.

\section{Discussion}
\label{section:discussion}

This study investigated how \emph{positive and negative media framing} modulates the influence of perceived risks and perceived benefits on the overall attributed value of HRC in a production context. 
Hereto, participants representative for the German working population evaluated short newspaper articles which were framed either as positively or negatively and focussed one of the three themes safety, autonomy, and job security.

The framing itself was highly effective, as indicated by the significant differences in the manipulation check: the newspaper articles were evaluated markedly different depending on the frame. 
Interestingly, our results suggest that the effect of framing is stronger at the affective rather than the cognitive level. 
While the items addressing the cognitive level were also significantly different between both frames, the effect was weaker compared to the affective dimension.

As expected in line with framing research, the type of framing affected the absolute levels of perceived risks, perceived benefits, and overall attributed value of HRC, queried after but independent of the framing intervention.
The findings were in line with our hypotheses (and our registration) and the results suggest that a positive frame yields higher evaluations for all three target dimensions compared to the negative one.
Thus, if HRC is presented in positive terms, it is---independent of the specific context---perceived as safer, more beneficial, and more valuable.
Hence, our study corroborates findings from prior work suggesting that media framing affects perceptions and attitudes \parencite{RoskosEwoldsen2007}.

Consistent with prior work, both perceived risks and perceived benefits were robust independent predictors in shaping the attribute value of cobots in both framing conditions.
Perceived risks \parencite{vanWynsberghe2012} and perceived benefits \parencite{deGraaf2015} have been highlighted as central factors in earlier research.
However, the strengths of the individual predictors varied across frames:
With negative framing, both perceived risks ($\beta \approx -.69$) and benefits ($\beta \approx \approx .63$) had stronger effects than with positive framing, where risks ($\beta \approx -.52$) and benefits ($\beta \approx .45$) showed slightly smaller effects.
Overall, while risks and benefits consistently emerged as strong predictors, their relative weight shifts from positive to negative framing.
In the positive frame, risks and benefits have similar weight, but although the absolute weights are slightly lower in the negative frame, the relative weight of the benefits becomes even less important. 

While the effect of framing on the \emph{absolute evaluations} is not particularly surprising, our findings further suggest that media framing alters \emph{how people weigh} risks and benefits when forming their overall value judgments.
The change manifests in two observations:
On the one hand, although both conditions showed a strong model fit (with the positive framing explaining more variance in attributed value ($R^2 = .715$) compared to the negative ($R^2 = .583$)), the interaction patterns between risk and benefit perceptions differed.
Participants’ value attributions towards HRC were more consistently predicted by perceived risks and benefits when messages were framed positively and emphasized potential gains rather than losses.
On the other hand, positive framing yielded a purely additive risk and benefit tradeoff for forming value judgements, whereas negative framing revealed a negative interaction between perceived risks and perceived benefits.
When outcomes are framed positively, people seem to treat risk and benefit as separable, independent factors.
This suggests that they evaluate them in a compensatory way, where high benefit can offset high risk, and vice versa.
In contrast, under negative framing, people do not simply add risks and benefits. 
The presence of potential losses dampens the influence of perceived benefits and triggers nonlinear evaluations.
As a result, when an option is perceived as unsafe, its overall perceived value decreases, even if the perceived benefits are substantial.

This relates to both prospect theory \parencite{Kahneman1979} and, more general, the Dual Process Model of cognition \parencite{Petty1986,Evans2013}.
The former posits that individuals evaluate outcomes relative to a reference point and are generally more sensitive to potential losses than equivalent gains.
The latter suggests two distinct modes of information processing:
a fast, intuitive, affect-driven system (System 1), and a slower, deliberative, analytical system (System 2).
With positive framing, the message likely activated System 1 processing, leading to heuristic-based evaluations that rely more directly on salient positive cues such as risks and benefits.
This is consistent with prior research showing that positively framed messages are more likely to elicit positive affect and intuitive trust \parencite{Rothman1997}.
In such conditions, individuals may independently (without interaction effects) and heuristically evaluate each attribute as contributing to the overall value of HRC without cognitively integrating or balancing them against each other.
Consequently, the relationships between perceived risks, benefits, and value appear stronger and more stable under positive framing, increasing the proportion of explained variance in the regression model.
Conversely, negatively framed messages may engage more elaborate, systematic processing (System 2), which involves greater cognitive effort and critical analysis \parencite{Evans2013}.
Participants exposed to the negatively framed messages may have weighed their perceived risks and benefits more critically.
This can introduce variability in how individuals integrate these cues, leading to less consistent and more complex evaluative patterns; ultimately reducing the model’s explanatory power.

We also conducted an exploratory analysis to examine how individual differences interact with message framing in shaping perceptions of risks and benefits in HRC.
The results suggest that neither age nor gender significantly influenced perceived risk of HRC.
However, gender had a small but statistically significant effect on perceived benefits when using positive framing, while age showed a similar effect with negative framing.
Beyond demographic factors and across both positive and negative frames, trust in robots consistently contributed to both perceived risks and benefits.
We attribute the relatively modest effects of demographic variables to the study’s design and would characterize the absence of effects as an experimental artefact.
While we prioritized the investigation of the framing effects over fine-grained measurement of individual differences in HRC perception, and studies focussing on the later suggest effects of gender and age on robot perception (cf. \textcite{Biermann2021,Broehl2019}).
Nevertheless, general trust in robots and automated systems appears to be a key mediator in shaping perceptions of both risks and benefits in HRC, which aligns with prior work \parencite{Broehl2019}.

These findings have practical implications for the design of HRC workplaces and communication of HRC strategies.
Messaging in industrial settings often oscillates between highlighting productivity gains and addressing safety concerns.
Our results suggest that when emphasizing potential losses (e.g., falling behind competitors without robots), communicators should be mindful of message saturation: layering too many negative cues may not yield proportional gains in perceived value.
In contrast, positively framed communications (e.g., emphasizing efficiency or innovation gains) may benefit from highlighting multiple positive dimensions, such as safety and utility, because their effects are more likely to be additive.

Methodologically, responsible research and innovation, user-centered and participatory implementation in Industry 4.0 and beyond often relies on acceptance models; either derivates of \textcite{Davis1989} Technology Acceptance Model (TAM) or the Value Based Adoption Model \parencite{Kim2007}.
These typically assume a stable function of risk, benefit, or other factors in driving evaluations.
Our findings challenge that assumption and show that contextual framing alters not only the absolute attitudes but the structure of the evaluation itself.
This suggests the need for dynamic models that incorporate situational or media-based moderators.

\section{Limitations and Future Research}\label{limitations}

While the current study provides robust evidence of framing effects in HRC evaluation, several limitations warrant attention.

First,
the study builds on a paid participant pool, so the participants' motivation to consciously respond to the survey question varies.
Nevertheless, we increased the quality of the data by checking attention items and removing speeders.
The latter is considered a sufficient quality control measure with large online panels \parencite{Leiner2019}.
An indicator for the achieved quality is the high reliability of the scales, despite negatively worded items.
Also, the results build on a sample from Germany, therefore generalizability or the influence of cultural effects need to be addressed.
Studies on the perception of robots suggest that cultural influences shouldn't be neglected as, for example, \textcite{Mims2010} found differences in robot perception between participants from Japan, Europe, and the USA.
We therefore suggest an international replication.

Second,
we used newspaper articles as framing stimuli.
Although this is a common medium in framing research, framing occurs across various media types, including videos, films, and interactive content and studies suggest an influence of the medium on perception \parencite{Zhang2009}.
While we assume that media framing is also effective in other formats, it remains unclear to what extent the medium itself modulates the formation of the risk-benefit tradeoffs.
More importantly, our study employed emphasis framing, highlighting the potential positive or negative influences of HRC.
Future research should complement this work by using equivalence framing, which contrasts outcomes presented as logically identical but differently valenced (e.g., gains vs. equivalent losses; \parencite{Cacciatore2015}).
Further, collaborative and social robots are increasingly used outside the production domain.
In this regard, we suggest studying the perception of robots and the influence of media framing in alternative contexts, such as healthcare and nursing (as we did in \parencite{Biermann2021} with different trust attributions between contexts).
A further intriguing question concerns the broader implications of media framing:
to what extent does it shape our understanding through repeated, complex, or subtle messages, such as those we encounter over time in news media or on social media?
Unlike isolated, one-time exposures, these long-term narratives can accumulate, subtly yet powerfully influencing public opinion and attitudes.

Third,
all data are based on perceived measures rather than behavioural outcomes, which may may be tainted by common method bias and limit causal inference.
Additionally, individual differences, such as self-efficacy in interacting with technology, influenced the perception of HRC.
While we have queried these as explanatory variables and checked the quality of the data, we have not specifically controlled for them in the sample composition and they could moderate the observed effects and should be explored in future work.
Finally, replicating this design in applied settings, such as virtual simulations or real production lines would enhance ecological validity and inform implementation strategies.

\section{Conclusion}
\label{section:conclusion}

This study examined the influence of media framing, delivered via newspaper articles, on public perceptions of HRC in industrial production.
Using a representative sample of the German population, we assessed the effects of framing on perceived risks, perceived benefits, and overall attributed value. 
Our findings indicate that media framing shapes both 1) the absolute evaluations of HRC and 2) the tradeoffs between perceived risks and benefits.

These results have several implications for the communication and acceptance of emerging technologies.
Framing not only alters the absolute magnitude of evaluations but also the cognitive architecture by which individuals weigh tradeoffs, shifting between additive (compensatory) and interactive (non-compensatory) judgment strategies.
In negatively framed contexts, risk often functions as a \enquote{dealbreaker}, with high benefits insufficient to offset perceived risks.
This suggests that in industries such as production, healthcare, aviation, food safety, or nuclear energy, the salience of potential negatives can diminish the persuasive power of benefits.
In such contexts, benefits should be framed as integral to safety rather than treated as separate dimensions.

As future research directions, our results highlight the context-sensitivity of value-based decision-making, challenging the established but static models of technology acceptance.
For policymakers, communicators, and other stakeholders, recognizing how media framing shapes public evaluation is essential for designing interventions that foster informed, balanced, and just responses to technological innovation.

\appendix
\clearpage
\setcounter{table}{0}
\renewcommand{\thetable}{\Alph{section}.\arabic{table}}

\section*{Acknowledgements}\label{acknowledgements}

We thank all participants of the survey for their commitment. We are also grateful to Hannah B. and Sarah H. for their earlier, yet unpublished, study that strongly informed this work. In addition, we thank Kjell F. and Julian H. for their constructive comments and careful corrections.
Funded by the Deutsche Forschungsgemeinschaft (DFG, German Research Foundation) under Germany's Excellence Strategy -- EXC-2023 Internet of Production -- 390621612.


%

%
%

\section{Additional Tables}

\begin{sidewaystable}[h!]
    \centering
    \begin{tabular}{lcp{10cm}cccc}
    \toprule
& & & \multicolumn{2}{c}{Loss Frame} & \multicolumn{2}{c}{Gain Frame} \\
\cmidrule(lr){4-5} \cmidrule(lr){6-7}
    \textbf{Dimension} & \textbf{Cronbach's $\alpha$} &	\textbf{Item} (P = Positive, N = Negative) & \textbf{M} & \textbf{SD} & \textbf{M} & \textbf{SD}\\ \midrule
\multirow{3}{*}{Perceived Risks} & \multirow{3}{*}{.725}	&	I have confidence that collaborative robots are safe. (N)	&	4.20	&	1.30	&	4.44	& 1.42	\\
&&	Collaborative robots minimise risks and ensure safety. (N) 	&	4.23	&	1.34 &	4.62		&	1.42 \\
&&	    Interaction with collaborative robots harbours significant risks. (P)	&	3.37	&	1.29 &	3.01&	1.34		\\ \midrule 
\multirow{3}{*}{Perceived Benefits} & \multirow{3}{*}{.722} 	& Collaborative robots increase efficiency and productivity. (P)	&	4.75&	1.37&	4.83	&	1.47\\
&& Interaction with collaborative robots provides valuable support and reduces the workload. (P) 	&	4.51	&	1.31&	4.75&	1.41	\\
&& Collaborative robots offer hardly any advantages compared to conventional tools. (N)	&	2.95 &	1.34 &	2.91 &	1.43 \\ \midrule
\multirow{3}{*}{Overall Value} & \multirow{3}{*}{.779}	&  Collaborative robots are a positive addition to many workplaces. (P) 	& 4.47	& 1.39 &		4.64& 1.45	\\
&& I feel optimistic about the increasing use of collaborative robots. (P) 		& 	4.11& 1.36	& 4.28	& 1.45	\\
&& I am sceptical about the benefits of collaborative robots in practice. (N) 	& 	3.34& 1.49	&	3.12&	1.51 \\
  	\bottomrule
    \end{tabular}
    \caption{Evaluation of the perceived risks, benefits, and attributed value of collaborative robots in production with item and scale characteristics (translated from the german original, scale from 1 (\emph{no agreement}) to 7 (\emph{full agreement})). Permission to use the items is granted given a proper citation.}
    \label{tab:robotevaluations}
\end{sidewaystable}

\begin{table}[ht]
\centering
\begin{threeparttable}
\caption{\label{tab:regressioncombined}Regression Results with Three-Way Interaction between Framing (loss vs. gain frame), and perceived benefits and risks on overall attributed value.}
\begin{tabular}{lrrrr}
\toprule
\textbf{Predictor} & \textbf{Std. $\beta$} & \textbf{Std. Error} & \textbf{$t$} & \textbf{$p$} \\
\midrule
(Intercept)                      & -0.091 & 0.044 & -2.054 & 0.040$^{*}$ \\
Perceived Risks                  & 0.687  & 0.085 & 8.097  & $<.001^{***}$ \\
Perceived Benefits	             & 0.630  & 0.076 & 8.308  & $<.001^{***}$ \\
Framing 			             & 0.082  & 0.056 & 1.463  & 0.144 \\
Risks $\times$ Benefits         & -0.324 & 0.117 & -2.771 & 0.006$^{**}$ \\
Risks $\times$ Framing          & -0.169 & 0.108 & -1.560 & 0.119 \\
Benefits $\times$ Framing        & -0.184 & 0.099 & -1.859 & 0.063$^{.}$ \\
Risks $\times$ Benefit $\times$ Framing & 0.330  & 0.144 & 2.282  & 0.023$^{*}$ \\
\midrule
\multicolumn{5}{l}{$F(7, 1143) = 311.4$,\quad $p < .001$, $R^2=.656$ ($R_{\text{adj.}}^2=.653$)} \\
\bottomrule
\end{tabular}
\begin{tablenotes}
\small
\item \textit{Note:} $^{***}p<0.001$, $^{**}p<0.01$, $^{*}p<0.05$, $^{.}p<0.1$
\end{tablenotes}
\end{threeparttable}
\end{table}

\begin{sidewaystable}[htbp]
\centering
\begin{tabular}{llp{0.8\textwidth}}
\toprule
\textbf{Framing} & \textbf{Context} & \textbf{Scenario text} shown as newspaper article (same image for each scenario) \\
\midrule

\multirow{3}{*}{\textbf{Positive}} 
& Employment & A study concludes that fears of job losses due to the use of robots and advancing digitalization are unfounded. Another study suggests that new technologies create more jobs than they eliminate. Sectors such as energy supply (30\%) and automotive manufacturing (25\%) are expected to experience significant employment growth. While robots may displace jobs initially, they ultimately create far more additional and higher-quality employment opportunities. \\
& Autonomy & Robots work quickly and accurately, are reliable, and cooperative. Autonomous robots are revolutionizing industries---to the benefit of workers. As intelligent tools, they offer more autonomy in daily work. Those learning a trade today no longer need to endure repetitive tasks indefinitely. Robots handle monotonous and repetitive tasks efficiently and effectively, allowing humans to focus more on creative and individualized responsibilities. \\
& Safety & Market research has shown that assembly robots can drastically reduce physical strain and accident risks. This is because robots take on dangerous tasks, enabling human workers to monitor from a safe distance. According to the study, severe injuries could decrease by 30\%, improving employee safety and reducing workplace absences. \\
\midrule

\multirow{3}{*}{\textbf{Negative}} 
& Employment & A study concludes that fears of job losses due to the use of robots and advancing digitalization are well-founded. Those entering the workforce today must anticipate that within a few years, a robot may perform their tasks. Many employees should prepare for the likelihood that robots will take over their jobs in the foreseeable future. High-risk sectors for job losses include the chemical industry (-30\%) and automotive manufacturing (-25\%). \\
& Autonomy & However, the future of industry shaped by autonomous robots could be detrimental to workers. The increasing use of robots in production risks undermining employees’ autonomy. Rigid programming allows robots to interfere with workers’ decision-making, control processes and work speed, and excessively monitor behavior. In Industry 4.0, humans risk being reduced to insignificant and interchangeable objects. \\
& Safety & Additionally, market research indicates that assembly robots could drastically increase physical strain and accident risks. This is because humans often need to operate in close proximity to robots to manage operations and supply components. According to the study, severe injuries could increase by 30\%, raising risks for employees and leading to higher rates of workplace absences. \\
\bottomrule
\end{tabular}
\caption{Positive and negative framing scenarios across employment, autonomy, and safety contexts.}
\label{tab:framing-scenarios}
\end{sidewaystable}

\clearpage

\printbibliography

\end{document}